\documentclass[letterpaper,english,pra, twocolumn,showpacs,superscriptaddress]{revtex4-1}

\usepackage[T1]{fontenc}
\usepackage{color}
\usepackage{babel}
\usepackage{pifont}
\usepackage{amsmath}
\usepackage{amssymb}
\usepackage{graphicx}
\usepackage[unicode=true,pdfusetitle,
 bookmarks=true,bookmarksnumbered=false,bookmarksopen=false,
 breaklinks=false,pdfborder={0 0 0},pdfborderstyle={},backref=false,colorlinks=true]
 {hyperref}
\hypersetup{
 urlcolor=blue, citecolor=blue}

\makeatletter

\pdfpageheight\paperheight
\pdfpagewidth\paperwidth

\usepackage{xcolor}

\AtBeginDocument{
  
}

\makeatother

\begin{document}

\title{The production rate of the system-bath mutual information}

\author{Sheng-Wen Li}

\affiliation{Texas A\&M University, College Station, TX 77843}

\affiliation{Baylor University, Waco, TX 76798}

\begin{abstract}
When an open system is contacted with several thermal baths, the entropy
produced by the irreversible processes ($dS_{\mathrm{i}}=dS-\sum_{\alpha}\text{\dj}Q_{\alpha}/T_{\alpha}$)
keeps increasing, and this entropy production rate is always non-negative.
But when the system is contacted with some non-thermal baths containing
quantum coherence or squeezing, this entropy production formula does
not apply. In this paper, we study the increasing rate of the mutual
information between the open system and its environment. When the
baths are canonical thermal ones, we prove that this mutual information
production rate could exactly return to the previous entropy production
rate. Further, we study an example of a single boson mode contacted
with multiple squeezed thermal baths, where the conventional entropy
production rate does not apply, and we find that this mutual information
production rate still keeps non-negative, which means the monotonic
increasing of the correlation between the system and its environment.
\end{abstract}
\maketitle

\section{Introduction }

The entropy change of a system can be considered to come from two
origins, i.e., $dS=dS_{\mathrm{e}}+dS_{\mathrm{i}}$ \cite{de_groot_non-equilibrium_1962,nicolis_self-organization_1977,reichl_modern_2009,kondepudi_modern_2014},
where $dS_{\mathrm{e}}$ comes from the exchange with external sources,
and it could be either positive or negative; $dS_{\mathrm{i}}$ is
the entropy change due to the irreversible processes. Then the 2nd
law is simply stated as  $dS_{\mathrm{i}}/dt\ge0$, which means the
entropy produced by the irreversible processes always increases, and
$R_{\mathrm{ep}}:=dS_{\mathrm{i}}/dt$ is called the \emph{entropy
production rate} (EPr).

\begin{figure}
\includegraphics[width=0.38\columnwidth]{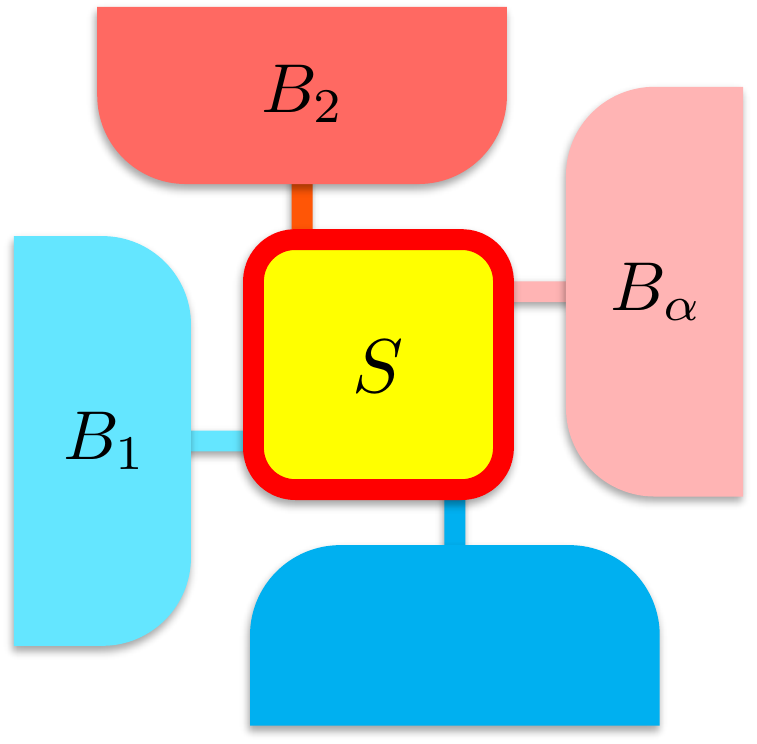}

\caption{(Color online) Demonstration for an open quantum system ($S$) interacting
with its environment composed of multiple baths ($B_{\alpha}$). The
baths are independent from each other, and do not have to be canonical
thermal states.}
\label{fig-mBaths}
\end{figure}

When the system is contacted with a thermal bath with temperature
$T$, we have $dS_{\mathrm{e}}=\text{\text{\dj}}Q/T$ (hereafter we
refer it as the \emph{thermal entropy}), where $\text{\dj}Q$ is the
heat flowing into the system. If we have multiple independent thermal
baths with different temperatures $T_{\alpha}$ (Fig.\,\ref{fig-mBaths}),
the EPr becomes 
\begin{equation}
\frac{dS_{\mathrm{i}}}{dt}=\frac{dS}{dt}-\sum_{\alpha}\frac{1}{T_{\alpha}}\frac{dQ_{\alpha}}{dt}:=R_{\mathrm{ep}},\label{eq:R-EPr}
\end{equation}
 where $\text{\dj}Q_{\alpha}$ is the heat coming from bath-$\alpha$
\cite{de_groot_non-equilibrium_1962,kondepudi_modern_2014}. 

Further, when an open quantum system is weakly coupled with the multiple
thermal baths, usually its dynamics can be described by the following
Lindblad (GKSL) equation \cite{gorini_completely_1976,lindblad_generators_1976},
\begin{equation}
\dot{\rho}=i[\rho,\hat{H}_{S}]+\sum_{\alpha}{\cal L}_{\alpha}[\rho].
\end{equation}
where ${\cal L}_{\alpha}[\rho]$ describes the dissipation due to
bath-$\alpha$. Utilizing $\dot{S}[\rho]=-\mathrm{tr}[\dot{\rho}\ln\rho]$
and $\dot{Q}_{\alpha}=\mathrm{tr}\big[\hat{H}_{S}\cdot{\cal L}_{\alpha}[\rho]\big]$,
the EPr (\ref{eq:R-EPr}) can be rewritten as the following Spohn
formula (denoted as $R_{\mathrm{Sp}}$ hereafter) \cite{spohn_entropy_1978,spohn_irreversible_1978,alicki_quantum_1979,boukobza_three-level_2007,kosloff_quantum_2013, *kosloff_quantum_2016,cai_entropy_2014} 
\begin{equation}
R_{\mathrm{ep}}=\sum_{\alpha}\mathrm{tr}\big[(\ln\rho_{\mathrm{ss}}^{(\alpha)}-\ln\rho)\cdot{\cal L}_{\alpha}[\rho]\big]:=R_{\mathrm{Sp}}.\label{eq:Spohn}
\end{equation}
Here we call $\rho_{\mathrm{ss}}^{(\alpha)}=Z_{\alpha}^{-1}\exp[-\hat{H}_{S}/T_{\alpha}]$
the \emph{partial steady state} associated with bath-$\alpha$, satisfying
${\cal L}_{\alpha}[\rho_{\mathrm{ss}}^{(\alpha)}]=0$. It can be proved
that $R_{\mathrm{Sp}}\ge0$, which means the irreversible entropy
production keeps increasing (see the proof in Appendix \ref{apx:Proof}
or Ref.\,\cite{spohn_entropy_1978,spohn_irreversible_1978}).

However, in the above discussion, the thermal entropy $dS_{\mathrm{e}}=\text{\dj}Q/T$
only applies for canonical thermal baths. If the bath is some non-canonical
state containing quantum coherence or squeezing \cite{scully_extracting_2003,rosnagel_nanoscale_2014,manzano_entropy_2016},
the temperature is not well defined, thus it is no more proper to
use $\text{\dj}Q/T$ for $dS_{\mathrm{e}}$ \cite{gardas_thermodynamic_2015},
and the relations $R_{\mathrm{ep}}=R_{\mathrm{Sp}}$ or $R_{\mathrm{ep}}\ge0$
no longer hold either.

Therefore, for such non-thermal baths, the conventional thermodynamic
description of the EPr does not apply. And it is believed that corrections
of some work \cite{quan_quantum_2005,quan_quantum-classical_2006,gelbwaser-klimovsky_heat-machine_2014},
or excess heat \cite{gardas_thermodynamic_2015} should be considered
in these baths.

Here we replace the thermal entropy term $-\dot{Q}_{\alpha}/T_{\alpha}$
by the von Neumann entropy of bath-$\alpha$, $\dot{S}_{B\alpha}=-\mathrm{tr}[\dot{\rho}_{B\alpha}\ln\rho_{B\alpha}]$.
Further, we assume the multiple baths are independent from each other,
thus it leads to $\sum_{\alpha}\dot{S}_{B\alpha}=\dot{S}_{B}$. Then
this generalization becomes
\begin{equation}
R_{{\cal I}}=\frac{dS_{S}}{dt}+\frac{dS_{B}}{dt}=\frac{d}{dt}(S_{S}+S_{B}-S_{SB})=\frac{d}{dt}{\cal I}_{SB}.\label{eq:R_EP_ISB}
\end{equation}
Here $\dot{S}_{SB}=0$ since the total system $S+B$ evolves unitarily
\footnote{$\dot{S}[\rho_{SB}]=-\mathrm{tr}[\dot{\rho}_{SB}\ln\rho_{SB}]=-i\,\mathrm{tr}[\rho_{SB}\hat{{\cal H}}\cdot\ln\rho_{SB}-\hat{{\cal H}}\rho_{SB}\cdot\ln\rho_{SB}]=0$,
where $\hat{{\cal H}}$ is the Hamiltonian of the total $S+B$ system.}, and ${\cal I}_{SB}:=S_{S}+S_{B}-S_{SB}$ is just the mutual information
between the system and its environment, which measures their correlation
\cite{nielsen_quantum_2000,esposito_entropy_2010,pucci_entropy_2013,parrondo_thermodynamics_2015,alipour_correlations_2016,strasberg_quantum_2017}.
Therefore, we call $R_{{\cal I}}$ the \emph{mutual information production
rate }(MIPr).

$R_{{\cal I}}$ has a clear physical meaning: a positive $R_{{\cal I}}$
indicates the correlation between the system and its environment is
increasing. In the following, we are going to show that, indeed this
MIPr (\ref{eq:R_EP_ISB}) has a quite close connection with the previous
EPr (\ref{eq:R-EPr}). When the bath of the open system are thermal
ones, we can prove that this MIPr could exactly return to the conventional
thermodynamic description of the EPr in the weak coupling limit, namely,
$R_{{\cal I}}=R_{\mathrm{ep}}$. That means, for thermal bath, the
conventional entropy production can be equivalently interpreted as
the mutual information production, and the 2nd law statement $(R_{{\cal I}}=)R_{\mathrm{ep}}\ge0$
can be also understood as the system-bath correlation always keeps
increasing.

Further, we will study an example of a single boson contacted with
multiple squeezed thermal baths. In this case, the conventional EPr
does not apply. We calculate the MIPr under the weak coupling limit
and Markovian approximation, and we find that it exactly equals to
the Spohn formula for non-thermal baths, thus we can prove $R_{{\cal I}}\ge0$,
which means the monotonic increasing of the system-bath correlation
also holds in this squeezed bath example. 

\section{Mutual information production in thermal baths}

Now we first consider the system is coupled with several thermal baths.
In this case, the initial state of bath-$\alpha$ is $\rho_{B\alpha}(0)={\cal Z}_{\alpha}^{-1}\exp[-\hat{H}_{B\alpha}/T_{\alpha}]$.
Assuming $\rho_{B\alpha}(t)$ does not change too much during evolution
\cite{breuer_theory_2002,scully_quantum_1997,li_non-markovianity_2016},
we have $\ln\rho_{B\alpha}(t)=\ln[\rho_{B\alpha}(0)+\delta\rho_{t}]\simeq\ln[\rho_{B\alpha}(0)]+o(\delta\rho_{t})$,
thus the entropy change of bath-$\alpha$ is 
\begin{align}
\dot{S}_{B\alpha} & =-\mathrm{tr}[\dot{\rho}_{B\alpha}(t)\ln\rho_{B\alpha}(t)]\simeq-\mathrm{tr}[\dot{\rho}_{B\alpha}(t)\cdot\ln\frac{e^{-\frac{\hat{H}_{B\alpha}}{T_{\alpha}}}}{{\cal Z}_{\alpha}}]\nonumber \\
 & =\frac{1}{T_{\alpha}}\frac{d}{dt}\langle\hat{H}_{B\alpha}\rangle\simeq-\frac{\dot{Q}_{\alpha}}{T_{\alpha}}.\label{eq:S_B-th}
\end{align}
Here $-\frac{d}{dt}\langle\hat{H}_{B\alpha}\rangle$ is the energy
loss of bath-$\alpha$, while $\dot{Q}_{\alpha}$ is the energy gain
of the system from bath-$\alpha$, and they equal to each other in
weak coupling limit. Assuming the baths are independent from each
other, $\rho_{B}(t)\simeq\prod_{\alpha}\rho_{B\alpha}(t)$, the MIPr
becomes
\begin{equation}
R_{{\cal I}}=\dot{S}_{S}+\sum_{\alpha}\dot{S}_{B\alpha}=\dot{S}_{S}-\sum_{\alpha}\frac{\dot{Q}_{\alpha}}{T_{\alpha}}=R_{\mathrm{ep}}.
\end{equation}
Therefore, for thermal baths, the MIPr (\ref{eq:R_EP_ISB}) equals
to the conventional thermodynamic description of the EPr (\ref{eq:R-EPr}).

Thus, the 2nd law statement $R_{\mathrm{ep}}\ge0$ is equivalent as
$R_{{\cal I}}\ge0$, which means the mutual information between the
system and its environment keeps increasing monotonically. This can
be understood as an equivalent statement for the entropy production
when the baths are canonical thermal ones. We notice that this equivalence
was also shown in the ``correlation entropy'' approach \cite{esposito_entropy_2010,pucci_entropy_2013,strasberg_quantum_2017}.

\section{Mutual information production in Squeezed baths}

Now we study an example of a single boson mode interacting with multiple
squeezed thermal baths \cite{rosnagel_nanoscale_2014,manzano_entropy_2016,kosloff_quantum_2016}.
In this case, the thermal entropy $dS_{\mathrm{e}}=\text{\dj}Q/T$
cannot be used, and neither does the EPr\,(\ref{eq:R-EPr}). Here
we calculate the MIPr (\ref{eq:R_EP_ISB}), and we will prove it just
equals to the Spohn formula for non-thermal baths, and thus could
still keep non-negative, $R_{{\cal I}}\ge0$.

\subsection{Master equation and Spohn formula}

The Hamiltonian of the single boson mode and the bosonic bath are
$\hat{H}_{S}=\Omega\,\hat{a}^{\dagger}\hat{a}$, $\hat{H}_{B}=\sum_{\alpha}\hat{H}_{B\alpha}$
and $\hat{H}_{B\alpha}=\sum_{k}\omega_{\alpha k}\,\hat{b}_{\alpha k}^{\dagger}\hat{b}_{\alpha k}$,
and they interact through $\hat{V}_{SB}=\sum_{\alpha}\hat{a}^{\dagger}\hat{B}_{\alpha}+\hat{a}\hat{B}_{\alpha}^{\dagger}$.
Here $\hat{B}_{\alpha}=\sum g_{\alpha k}\hat{b}_{\alpha k}$ is the
operator of bath-$\alpha$, and the initial states of the baths are
squeezed thermal ones (hereafter all the density matrices are written
in the interaction picture),
\begin{gather}
\rho_{B\alpha}^{0}=\frac{1}{{\cal Z}_{\alpha}}e^{-\beta_{\alpha}\,{\cal S}_{\alpha}\hat{H}_{B\alpha}{\cal S}_{\alpha}^{\dagger}},\quad\beta_{\alpha}:=T_{\alpha}^{-1},\label{eq:bath-Sq}\\
{\cal S}_{\alpha}:=\prod_{k}\exp[\frac{1}{2}\lambda_{\alpha k}^{*}\hat{b}_{\alpha k}^{2}-\mathbf{h.c.}],\quad\lambda_{\alpha k}=r_{\alpha k}e^{-i\theta_{\alpha k}}.\nonumber 
\end{gather}
Here ${\cal S}_{\alpha}$ is the squeezing operator for the boson
modes in bath-$\alpha$. With Born-Markovian approximation, we obtain
a master equation $\dot{\rho}=\sum_{\alpha}{\cal L}_{\alpha}[\rho]$
for the open system alone \cite{breuer_theory_2002,walls_quantum_2008},
where
\begin{align*}
{\cal L}_{\alpha}[\rho] & =\frac{\gamma_{\alpha}}{2}\Big[\tilde{\mathfrak{n}}_{\alpha}\big(2\hat{a}^{\dagger}\rho\hat{a}-\{\hat{a}\hat{a}^{\dagger},\rho\}\big)+(\tilde{\mathfrak{n}}_{\alpha}+1)\big(2\hat{a}\rho\hat{a}^{\dagger}-\{\hat{a}^{\dagger}\hat{a},\rho\}\big)\\
 & -\tilde{\mathfrak{u}}_{\alpha}\big(2\hat{a}^{\dagger}\rho\hat{a}^{\dagger}-\{(\hat{a}^{\dagger})^{2},\rho\}\big)-\tilde{\mathfrak{u}}_{\alpha}^{*}\big(2\hat{a}\rho\hat{a}-\{\hat{a}^{2},\rho\}\big)\Big].
\end{align*}
The coupling spectrums of the squeezed bath-$\alpha$ are $J_{\alpha}(\omega):=2\pi\sum_{k}|g_{\alpha k}|^{2}\delta(\omega-\omega_{\alpha k})$
and $K_{\alpha}(\omega):=2\pi\sum_{k}g_{\alpha k}^{2}\delta(\omega-\omega_{\alpha k})$.
Without loss of generality, we omit the phase of $g_{\alpha k}$ and
thus $K_{\alpha}(\omega)=K_{\alpha}^{*}(\omega)=J_{\alpha}(\omega)$.
Here we denote $\gamma_{\alpha}:=J_{\alpha}(\Omega)=K_{\alpha}(\Omega)$,
and the parameters $\tilde{\mathfrak{n}}_{\alpha}:=\tilde{\mathsf{n}}_{\alpha}(\Omega)$,
$\tilde{\mathfrak{u}}_{\alpha}:=\tilde{\mathsf{u}}_{\alpha}(\Omega)$
are calculated from $\tilde{\mathsf{n}}_{\alpha}(\omega_{k}):=\mathrm{tr}[\rho_{B\alpha}^{0}\hat{b}_{\alpha k}^{\dagger}\hat{b}_{\alpha k}]$,
$\tilde{\mathsf{u}}_{\alpha}(\omega_{k}):=-\mathrm{tr}[\rho_{B\alpha}^{0}\hat{b}_{\alpha k}^{2}]$
(see Appendix \ref{apx:squeeze}). The master equation gives
\begin{gather}
\frac{d}{dt}\langle\tilde{a}(t)\rangle=-\sum_{\alpha}\frac{\gamma_{\alpha}}{2}\langle\tilde{a}\rangle,\quad\frac{d}{dt}\langle\tilde{a}^{2}\rangle=-\sum_{\alpha}\gamma_{\alpha}[\langle\tilde{a}^{2}\rangle-\tilde{\mathfrak{u}}_{\alpha}],\nonumber \\
\frac{d}{dt}\langle\tilde{a}^{\dagger}\tilde{a}\rangle=-\sum_{\alpha}\gamma_{\alpha}[\langle\tilde{n}_{a}\rangle-\tilde{\mathfrak{n}}_{\alpha}].\label{eq:dynamics}
\end{gather}
Here we denote $\hat{n}_{a}:=\hat{a}^{\dagger}\hat{a}$, and $\langle\tilde{o}(t)\rangle:=\mathrm{tr}[\rho\hat{o}(t)]$
gives variables in the rotating frame \footnote{Here $\rho$ is in
the interaction picture, but $\hat{o}$ is in the Schr{\"o}dinger
picture, thus we have { $\langle\hat{a}(t)\rangle=\langle\tilde{a}(t)\rangle e^{-i\Omega t}$}.
Here {$\langle\hat{o}(t)\rangle$} stands for observable expectations
which are independent of pictures, and {$\langle\tilde{o}(t)\rangle$}
are variables in the rotating frame, thus in Eq.\,(\ref{eq:dynamics}),
the dependence of the system frequency $\Omega$ is cancelled.}.

The partial steady states $\rho_{\mathrm{ss}}^{(\alpha)}$, which
satisfies ${\cal L}_{\alpha}[\rho_{\mathrm{ss}}^{(\alpha)}]=0$, are
now squeezed thermal states,
\begin{gather}
\rho_{\mathrm{ss}}^{(\alpha)}=\frac{1}{Z_{\alpha}}\exp[-\beta_{\alpha}\Omega\cdot\mathsf{S}_{\alpha}\hat{a}^{\dagger}\hat{a}\mathsf{S}_{\alpha}^{\dagger}],\\
\mathsf{S}_{\alpha}:=\exp[-(\frac{1}{2}\zeta_{\alpha}^{*}\hat{a}^{2}-\mathbf{h.c.})],\:\zeta_{\alpha}=\lambda_{\alpha k}\big|_{\omega_{k}=\Omega}:=r_{\alpha}e^{i\theta_{\alpha}}.\nonumber 
\end{gather}
Here $\mathsf{S}_{\alpha}$ is a squeezing operator for the system.
Although the baths are not thermal ones, we can still write down the
Spohn formula $R_{\mathrm{Sp}}=\sum_{\alpha}R_{\mathrm{Sp}}^{(\alpha)}$,
where 
\begin{align}
R_{\mathrm{Sp}}^{(\alpha)} & :=\mathrm{tr}\big[(\ln\rho_{\mathrm{ss}}^{(\alpha)}-\ln\rho)\cdot{\cal L}_{\alpha}[\rho]\big]\nonumber \\
 & :=\chi_{\alpha}-\mathrm{tr}\big[\ln\rho\cdot{\cal L}_{\alpha}[\rho]\big]\label{eq:R_S-X}
\end{align}
and we can prove $R_{\mathrm{Sp}}^{(\alpha)}\ge0$ and $R_{\mathrm{Sp}}\ge0$
hold also in this non-thermal case (Appendix \ref{apx:Proof}).

However, since the above Spohn formula $R_{\mathrm{Sp}}$ for non-thermal
baths no more comes from the thermodynamic EPr\,(\ref{eq:R-EPr}),
thus its physical meaning is unclear now. In the thermal case, the
1st term in $R_{\mathrm{Sp}}^{(\alpha)}$, $\chi_{\alpha}:=\mathrm{tr}\big[\ln\rho_{\mathrm{ss}}^{(\alpha)}\cdot{\cal L}_{\alpha}[\rho]\big]$,
gives the changing rate of the thermal entropy ($\chi_{\alpha}=-\dot{Q}_{\alpha}/T_{\alpha}$).
But for the squeezed case, it becomes
\begin{multline}
\chi_{\alpha}=\frac{\Omega}{T_{\alpha}}\cdot\gamma_{\alpha}\Big(\cosh2r_{\alpha}\cdot[\langle\tilde{n}_{a}(t)\rangle-\tilde{\mathfrak{n}}_{\alpha}]\\
-\frac{1}{2}\sinh2r_{\alpha}[e^{-i\theta_{\alpha}}(\langle\tilde{a}^{2}(t)\rangle-\tilde{\mathfrak{u}}_{\alpha})+\mathbf{h.c.}]\Big).\label{eq:sigma_B}
\end{multline}
It is difficult to tell the physical meaning of this quantity. In
the following, we will show that indeed Eq.\,(\ref{eq:sigma_B})
is just the changing rate of the von Neumann entropy of bath-$\alpha$,
i.e., $\chi_{\alpha}=\dot{S}_{B\alpha}$, and then Eq.\,(\ref{eq:R_S-X})
directly leads to
\begin{equation}
R_{\mathrm{Sp}}=\sum_{\alpha}R_{\mathrm{Sp}}^{(\alpha)}=\dot{S}_{S}+\sum_{\alpha}\dot{S}_{B\alpha}=R_{{\cal I}}.
\end{equation}

\subsection{Bath entropy dynamics}

Now we are going to calculate the entropy changing rate $\dot{S}_{B\alpha}$
of bath-$\alpha$ directly. To do this, we adopt the same trick as
the thermal case. Assuming the squeezed baths do not change too much
(interaction picture), the entropy of the bath evolves as 
\begin{align}
\frac{d}{dt}S[\rho_{B\alpha}(t)]\simeq- & \mathrm{tr}[\dot{\rho}_{B\alpha}(t)\cdot\ln\frac{\exp[-\beta_{\alpha}\,{\cal S}_{\alpha}\hat{H}_{B\alpha}{\cal S}_{\alpha}^{\dagger}]}{{\cal Z}_{\alpha}}]\nonumber \\
=\frac{d}{dt}\sum_{k}\frac{\omega_{\alpha k}}{T_{\alpha}}\Big( & \cosh2r_{\alpha k}\langle\tilde{b}_{\alpha k}^{\dagger}(t)\tilde{b}_{\alpha k}(t)\rangle\nonumber \\
+\frac{1}{2}\sinh & 2r_{\alpha k}[\langle\tilde{b}_{\alpha k}^{2}(t)\rangle e^{-i\theta_{\alpha k}}+\mathbf{h.c.}]\Big).\label{eq:Sq-evolu}
\end{align}
Thus, the calculation of the bath entropy is now reduced as calculating
the time derivative of the expectations of the bath operators like
$\langle\tilde{b}_{\alpha k}^{\dagger}(t)\tilde{b}_{\alpha k}(t)\rangle$
and $\langle\tilde{b}_{\alpha k}^{2}(t)\rangle$.

This can be done with the help of the Heisenberg equations, $\dot{\hat{b}}_{\alpha k}=-i\omega_{\alpha k}\hat{b}_{\alpha k}-ig_{\alpha k}^{*}\hat{a}$,
and $\dot{\hat{a}}=-i\Omega\hat{a}-i\sum_{\alpha}g_{\alpha k}\hat{b}_{\alpha k}$,
which lead to the quantum Langevin equation \cite{gardiner_quantum_2004,walls_quantum_2008,li_probing_2014}
\begin{equation}
\frac{d}{dt}\hat{a}=-i\Omega\hat{a}-\frac{1}{2}\Gamma\hat{a}-\hat{{\cal E}}(t).\label{eq:b(t)a(t)}
\end{equation}
Here $\Gamma:=\sum_{\alpha}\gamma_{\alpha}$ is the total decay rate,
and $\gamma_{\alpha}$ are the same as those in the master equation;
$\hat{{\cal E}}(t):=\sum_{\alpha}\hat{\xi}_{\alpha}(t)$ is the random
force, and $\hat{\xi}_{\alpha}(t):=i\sum_{k}g_{\alpha k}\hat{b}_{\alpha k}(0)e^{-i\omega_{\alpha k}t}$
is the contribution from bath-$\alpha$. Thus $\hat{a}(t)$ and $\hat{b}_{\alpha k}(t)$
evolve as 
\begin{gather}
\hat{a}(t)=\hat{a}(0)e^{-i\Omega t-\frac{\Gamma}{2}t}-\int_{0}^{t}ds\,e^{-i\Omega(t-s)-\frac{\Gamma}{2}(t-s)}\hat{{\cal E}}(s),\label{eq:a(t)}\\
\hat{b}_{\alpha k}(t)=\hat{b}_{\alpha k}(0)e^{-i\omega_{\alpha k}t}-ig_{\alpha k}^{*}\int_{0}^{t}ds\,e^{-i\omega_{\alpha k}(t-s)}\hat{a}(s).\nonumber 
\end{gather}

To further calculate the bath entropy change, now we are going to
show the following two relations hold in the weak coupling limit and
Markovian approximation:
\begin{align}
\frac{d}{dt}\sum_{k}\mathfrak{f}_{k}\langle\tilde{b}_{\alpha k}^{\dagger}\tilde{b}_{\alpha k}\rangle & \simeq\mathfrak{f}(\Omega)\cdot\gamma_{\alpha}[\langle\tilde{n}_{a}\rangle-\tilde{\mathfrak{n}}_{\alpha}],\label{eq:F-k}\\
\frac{d}{dt}\sum_{k}\mathfrak{h}_{k}\langle\tilde{b}_{\alpha k}^{2}\rangle+\mathbf{h.c.} & \simeq-\mathfrak{h}(\Omega)\cdot\gamma_{\alpha}[\langle\tilde{a}^{2}\rangle-\tilde{\mathfrak{u}}_{\alpha}]+\mathbf{h.c.},\nonumber 
\end{align}
where $\mathfrak{f}_{k}$ and $\mathfrak{h}_{k}$ are arbitrary coefficients
depending on $k$.

If we set $\mathfrak{f}_{k}=\frac{\omega_{\alpha k}}{T_{\alpha}}\cosh2r_{\alpha k}$,
$\mathfrak{h}_{k}=\frac{\omega_{\alpha k}}{2T_{\alpha}}\sinh2r_{\alpha k}e^{-i\theta_{\alpha k}}$,
and sum up the above two equations, then the left side simply gives
$\dot{S}_{B\alpha}$ {[}Eq.\,(\ref{eq:Sq-evolu}){]}; At the same
time, the right side is just equal to $\chi_{\alpha}$ {[}Eq.\,(\ref{eq:sigma_B}){]}.
Thus we can prove $\chi_{\alpha}=\dot{S}_{B\alpha}$, namely, the
term $\chi_{\alpha}=\mathrm{tr}\big[\ln\rho_{\mathrm{ss}}^{(\alpha)}\cdot{\cal L}_{\alpha}[\rho]\big]$
in the Spohn formula is just the changing rate of the von Neumann
entropy of bath-$\alpha$.

Besides, if we set $\mathfrak{f}_{k}=\omega_{\alpha k}$ and $\mathfrak{h}_{k}=0$,
the above relations lead to $\frac{d}{dt}\langle\hat{H}_{B\alpha}\rangle=\Omega\cdot\gamma_{\alpha}[\langle\tilde{n}_{a}\rangle-\tilde{\mathfrak{n}}_{\alpha}]=-\dot{Q}_{\alpha}$,
which means the energy loss of bath-$\alpha$ is equal to the energy
gain of the system from bath-$\alpha$ {[}as we utilized in the discussion
below Eq.\,(\ref{eq:S_B-th}){]}.

The calculation of Eq.\,(\ref{eq:F-k}) goes as follows
\begin{align}
 & \frac{d}{dt}\sum_{k}\mathfrak{f}_{k}\langle\tilde{b}_{\alpha k}^{\dagger}\tilde{b}_{\alpha k}\rangle=\sum_{k}\mathfrak{f}_{k}\cdot ig_{\alpha k}\langle\hat{a}^{\dagger}\hat{b}_{\alpha k}\rangle+\mathbf{h.c.}\nonumber \\
= & \sum_{k}\mathfrak{f}_{k}\cdot\Big[ig_{\alpha k}\langle\hat{a}^{\dagger}(t)\hat{b}_{\alpha k}(0)\rangle e^{-i\omega_{\alpha k}t}\nonumber \\
 & +|g_{\alpha k}|^{2}\int_{0}^{t}ds\,e^{-i\omega_{\alpha k}(t-s)}\langle\hat{a}^{\dagger}(t)\hat{a}(s)\rangle\Big]+\mathbf{h.c.}\label{eq:Part1}
\end{align}
 The 1st term in the bracket can be further calculated by substituting
$\hat{a}(t)$ {[}Eq.\,(\ref{eq:a(t)}){]},
\begin{align}
 & \sum_{k}\mathfrak{f}_{k}\cdot ig_{\alpha k}\langle\hat{a}^{\dagger}(t)\hat{b}_{\alpha k}(0)\rangle e^{-i\omega_{\alpha k}t}+\mathbf{h.c.}\\
= & -\sum_{k}\mathfrak{f}_{k}|g_{\alpha k}|^{2}\int_{0}^{t}ds\,e^{[i(\Omega-\omega_{k})-\frac{\Gamma}{2}](t-s)}\langle\hat{b}_{\alpha k}^{\dagger}(0)\hat{b}_{\alpha k}(0)\rangle+\mathbf{h.c.}\nonumber \\
= & -\int_{0}^{t}ds\big[\int_{0}^{\infty}\frac{d\omega}{2\pi}e^{[i(\Omega-\omega)-\frac{\Gamma}{2}](t-s)}J_{\alpha}(\omega)\mathfrak{f}(\omega)\tilde{\mathsf{n}}_{\alpha}(\omega)\big]+\mathbf{h.c.}\nonumber 
\end{align}
Assuming the frequency integral in the bracket gives a fast-decaying
function of $(t-s)$, we extend the time integral to $t\rightarrow\infty$
(Markovian approximation), and that gives
\begin{align}
 & -\int_{0}^{\infty}\frac{d\omega}{2\pi}[\int_{0}^{\infty}ds\,e^{i(\Omega-\omega)s-\frac{1}{2}\Gamma s}]J_{\alpha}(\omega)\mathfrak{f}(\omega)\tilde{\mathsf{n}}_{\alpha}(\omega)+\mathbf{h.c.}\nonumber \\
= & -\int_{0}^{\infty}\frac{d\omega}{2\pi}\,J_{\alpha}(\omega)\mathfrak{f}(\omega)\tilde{\mathsf{n}}_{\alpha}(\omega)\cdot\frac{\Gamma}{(\frac{\Gamma}{2})^{2}+(\omega-\Omega)^{2}}\nonumber \\
\simeq & -\mathfrak{f}(\Omega)\cdot\gamma_{\alpha}\tilde{\mathfrak{n}}_{\alpha}.\label{eq:part1-1}
\end{align}
The last line holds in the weak coupling limit $\Gamma\ll\Omega$
because the Lorentzian function in the integral approaches $2\pi\delta(\omega-\Omega)$.

To calculate the 2nd term of Eq.\,(\ref{eq:Part1}), we should notice
$\langle\hat{a}^{\dagger}(t)\hat{a}(s)\rangle=\langle\tilde{a}^{\dagger}(s)\tilde{a}(s)\rangle e^{(i\Omega-\frac{\Gamma}{2})(t-s)}$
holds for $t\ge s$ (quantum regression theorem \cite{breuer_theory_2002,gardiner_quantum_2004}).
Here $\langle\tilde{o}_{1}(t)\tilde{o}_{2}(s)\rangle$ is a correlation
function in the rotating frame, defined by $\langle\tilde{o}_{1}(t)\tilde{o}_{2}(s)\rangle=\mathrm{tr}[\hat{o}_{1}\,{\cal E}_{t-s}\hat{o}_{2}\,{\cal E}_{s}\rho(0)]$
for $t\ge s$ \cite{breuer_theory_2002}, where $\hat{o}_{1,2}$ are
operators in Schr\"odinger picture, and ${\cal E}_{t}$ is the evolution
operator solved from the above master equation in interaction picture,
and $\rho(t)={\cal E}_{t-s}\rho(s)$. Similarly, $\langle\hat{o}_{1}(t)\hat{o}_{2}(s)\rangle$
are correlation functions in the non-rotating frame. Thus the 2nd
term of Eq.\,(\ref{eq:Part1}) gives
\begin{align}
 & \sum_{k}\mathfrak{f}_{k}\cdot|g_{\alpha k}|^{2}\int_{0}^{t}ds\,e^{-i\omega_{\alpha k}(t-s)}\langle\hat{a}^{\dagger}(t)\hat{a}(s)\rangle+\mathbf{h.c.}\nonumber \\
\simeq & \int_{0}^{\infty}\frac{d\omega}{2\pi}\,\mathfrak{f}(\omega)J_{\alpha}(\omega)\cdot\langle\tilde{n}_{a}(t)\rangle\int_{0}^{\infty}ds\,e^{i(\Omega-\omega)s-\frac{\Gamma}{2}s}+\mathbf{h.c.}\nonumber \\
= & \langle\tilde{n}_{a}(t)\rangle\cdot\int_{0}^{\infty}\frac{d\omega}{2\pi}\,\mathfrak{f}(\omega)J_{\alpha}(\omega)\cdot\frac{\Gamma}{(\frac{\Gamma}{2})^{2}+(\omega-\Omega)^{2}}\nonumber \\
\simeq & \gamma_{\alpha}\cdot\mathfrak{f}(\Omega)\langle\tilde{n}_{a}(t)\rangle.\label{eq:part1-2}
\end{align}
Again we adopted the Markovian approximations as before, and $\langle\tilde{n}_{a}(s)\rangle$
is taken out of the integral directly.

Therefore, summing up Eqs.\,(\ref{eq:part1-1}, \ref{eq:part1-2}),
we obtain the 1st relation in Eq.\,(\ref{eq:F-k}). The 2nd relation
can be obtained through the similar way (see Appendix \ref{apx:squeeze}).
Then, by setting proper coefficients $\mathfrak{f}_{k}$ and $\mathfrak{h}_{k}$
in Eq.\,(\ref{eq:F-k}), we can prove $\chi_{\alpha}=\dot{S}_{B\alpha}$,
and further $R_{{\cal I}}=R_{\mathrm{Sp}}$. Since we can prove the
Spohn formula $R_{\mathrm{Sp}}\ge0$, the MIPr $R_{{\cal I}}$ also
keeps positive, which means the the system-bath mutual information,
or their correlation, still keeps increasing monotonically in this
non-thermal case.

\section{Summary }

In this paper, we study the production of the mutual information between
the system and its environment. We find that this MIPr\,(\ref{eq:R_EP_ISB})
has a close connection with the conventional thermodynamic description
of the EPr\,(\ref{eq:R-EPr}): when the baths of the open system
are canonical thermal ones, this MIPr could exactly return to the
previous EPr. Therefore, the 2nd law statement $R_{\mathrm{ep}}\ge0$
can be equivalently understood as saying the system-bath correlation
always keeps increasing.

Besides, we also study an example of a single boson mode contacted
with multiple squeezed thermal baths. In this case, the temperatures
of the baths are not well defined and the previous EPr does not apply.
We proved that the MIPr is still positive, which means the monotonic
increasing of the system-bath correlation also exists in this case.
Definitely it is worthful to study the MIPr in more non-thermal systems.

We remark that the proof for the positivity of the MIPr and the Spohn
formula relies on the fact the dynamics of the system can be well
described by a Markovian master equation with the Lindblad (GKSL)
form. If this is not fulfilled \cite{pucci_entropy_2013,sharma_landauer_2015,li_non-markovianity_2016,lampo_lindblad_2016},
the  positivity cannot be guaranteed.

Our study indicates it is the system-bath correlation that keeps increasing
monotonically although the total $S+B$ system evolves unitarily.
This idea is also consistent with some other fundamental studies on
thermodynamics, such as the local relaxation hypothesis \cite{cramer_exact_2008, *eisert_quantum_2015},
the entanglement based thermodynamics \cite{popescu_entanglement_2006, *goldstein_canonical_2006},
and the mutual information understanding of the Blackhole radiation
\cite{zhang_hidden_2009, *zhang_entropy_2011}.

\emph{Acknowledgement} \textendash{} The author appreciate much for
the helpful discussions with G. Agarwal, H. Dong, M. B. Kim, T. Peng,
M. O. Scully, A. Svidzinsky, D. Wang in Texas A\&M University, and
C. P. Sun in Beijing Computational Science Research Center. This study
is supported by Office of Naval Research (Award No. N00014-16-1-3054)
and Robert A. Welch Foundation (Grant No. A-1261).

\appendix
\begin{widetext}

\section{Proof for the positivity of Spohn formula \label{apx:Proof}}

Now we prove, the Spohn formula $R_{\mathrm{Sp}}$ is positive also
for non-thermal baths. Namely, for a Lindblad (GKSL) master equation
like \cite{gorini_completely_1976,lindblad_generators_1976}
\begin{gather}
\dot{\rho}=i[\rho,\,\hat{H}_{S}]+\sum_{\alpha}{\cal L}_{\alpha}[\rho]:=i[\rho,\,\hat{H}_{S}]+{\cal L}[\rho],\nonumber \\
{\cal L}_{\alpha}[\rho]=\sum_{n}V_{\alpha,n}\rho V_{\alpha,n}^{\dagger}-\frac{1}{2}\{V_{\alpha,n}^{\dagger}V_{\alpha,n},\,\rho\},
\end{gather}
we have 
\begin{equation}
R_{\mathrm{Sp}}^{(\alpha)}=\mathrm{tr}\left[{\cal L}_{\alpha}[\rho]\cdot(\ln\rho_{\mathrm{ss}}^{(\alpha)}-\ln\rho)\right]\ge0,\qquad R_{\mathrm{Sp}}=\sum_{\alpha}R_{\mathrm{Sp}}^{(\alpha)}\ge0,\label{eq:R_S}
\end{equation}
where $\rho_{\mathrm{ss}}^{(\alpha)}$ is the partial steady state
satisfying ${\cal L}_{\alpha}[\rho_{\mathrm{ss}}^{(\alpha)}]=0$.
The operator ${\cal L}_{\alpha}[\rho]$ describes the dissipation
to bath-$\alpha$, which does not have to be a thermal bath, and $\rho_{\mathrm{ss}}^{(\alpha)}$
is not necessarily a thermal state. 

Our proof follows from Ref.\,\cite{spohn_entropy_1978}, where a
single bath was concerned and the EPr was defined by the relative
entropy \cite{horowitz_equivalent_2014,manzano_entropy_2016},
\begin{equation}
\sigma=-\frac{d}{dt}S[\rho(t)\parallel\rho_{\mathrm{ss}}]=\mathrm{tr}\left[{\cal L}[\rho]\cdot(\ln\rho_{\mathrm{ss}}-\ln\rho)\right].
\end{equation}
Here $\rho_{\mathrm{ss}}$ is the steady state of the system satisfying
${\cal L}[\rho_{\mathrm{ss}}]=0$. This is equivalent with Eq.\,(\ref{eq:R_S})
when only one single bath is concerned. This EPr based on relative
entropy always gives $\sigma=0$ at the steady state, even for the
non-equilibrium steady state when there are multiple baths and usually
a steady non-equilibrium flux exists. But the EPr we used {[}Eq.\,(\ref{eq:R_S}){]}
will remain non-zero in this case, which means the irreversible entropy
is still being produced in the non-equilibrium steady state.

The proof for the positivity of Eq.\,(\ref{eq:R_S}) goes as follows.

\emph{Proof}: Since the master equation has the Lindblad (GKSL) form,
we obtain
\begin{equation}
\mathrm{tr}\left[{\cal L}_{\alpha}[\rho]\ln\rho\right]=\sum_{n}\mathrm{tr}\left[V_{\alpha,n}\rho V_{\alpha,n}^{\dagger}\ln\rho-V_{\alpha,n}^{\dagger}V_{\alpha,n}\rho\ln\rho\right].
\end{equation}

Now we need the Lieb theorem \cite{lieb_convex_1973}, namely, the
functional $f_{q}^{(V)}[\rho]=-\mathrm{tr}\left[\rho^{q}V\rho^{1-q}V^{\dagger}\right]$
is convex for $\forall\,0\le q\le1$, i.e.,
\begin{equation}
f_{q}^{(V)}[\lambda_{1}\rho_{1}+\lambda_{2}\rho_{2}]\le\lambda_{1}f_{q}^{(V)}[\rho_{1}]+\lambda_{2}f_{q}^{(V)}[\rho_{2}].
\end{equation}
At $q=0$, $f_{q=0}[\rho]=-\mathrm{tr}[V\rho V^{\dagger}]$ is an
linear map satisfying $f_{0}[\lambda_{1}\rho_{1}+\lambda_{2}\rho_{2}]=\lambda_{1}f_{0}[\rho_{1}]+\lambda_{2}f_{0}[\rho_{2}]$,
therefore, the derivative $\partial_{q}f_{q}^{(V)}:=\epsilon^{-1}[f_{q+\epsilon}^{(V)}-f_{q}^{(V)}]$
is also convex around $q=0$, which reads,
\begin{align}
\partial_{q}f_{q}^{(V)}[\rho]\Big|_{q=0} & =\mathrm{tr}\left[\rho^{q}V\rho^{1-q}\ln\rho V^{\dagger}-\rho^{q}\ln\rho V\rho^{1-q}V^{\dagger}\right]\Big|_{q=0}\nonumber \\
 & =\mathrm{tr}[V^{\dagger}V\rho\ln\rho-V\rho V^{\dagger}\ln\rho]:=-\mathrm{tr}\left[\mathbf{L}_{V}[\rho]\cdot\ln\rho\right].
\end{align}
Here we denoted $\hat{\mathbf{L}}_{V}[\rho]:=V\rho V^{\dagger}-\frac{1}{2}\{V^{\dagger}V,\,\rho\}$.
Thus, we obtain the following relation ($\lambda\ge0$),
\begin{align}
\partial_{q}f_{q}^{(V)}[\lambda\rho+(1-\lambda)\rho_{\mathrm{ss}}^{(\alpha)}] & =-\mathrm{tr}\left[\hat{\mathbf{L}}_{V}[\lambda\rho+(1-\lambda)\rho_{\mathrm{ss}}^{(\alpha)}]\cdot\ln[\lambda\rho+(1-\lambda)\rho_{\mathrm{ss}}^{(\alpha)}]\right]\nonumber \\
\le\lambda\cdot\partial_{q}f_{q}^{(V)}[\rho]+ & (1-\lambda)\cdot\partial_{q}f_{q}^{(V)}[\rho_{\mathrm{ss}}^{(\alpha)}]=-\lambda\mathrm{tr}\left[\hat{\mathbf{L}}_{V}[\rho]\cdot\ln\rho\right]-(1-\lambda)\mathrm{tr}\left[\hat{\mathbf{L}}_{V}[\rho_{\mathrm{ss}}^{(\alpha)}]\cdot\ln\rho_{\mathrm{ss}}^{(\alpha)}\right].
\end{align}

Since the Lindblad operator can be written as ${\cal L}_{\alpha}[\rho]=\sum_{n}\hat{\mathbf{L}}_{V_{\alpha,n}}[\rho]$,
from the above relation we obtain,
\begin{equation}
-\mathrm{tr}\left[{\cal L}_{\alpha}[\lambda\rho+(1-\lambda)\rho_{\mathrm{ss}}^{(\alpha)}]\cdot\ln[\lambda\rho+(1-\lambda)\rho_{\mathrm{ss}}^{(\alpha)}]\right]\le-\lambda\mathrm{tr}\left[{\cal L}_{\alpha}[\rho]\cdot\ln\rho\right]-(1-\lambda)\mathrm{tr}\left[{\cal L}_{\alpha}[\rho_{\mathrm{ss}}^{(\alpha)}]\cdot\ln\rho_{\mathrm{ss}}^{(\alpha)}\right].
\end{equation}
Here ${\cal L}_{\alpha}$ is a linear operator, thus, ${\cal L}_{\alpha}[\lambda\rho+(1-\lambda)\rho_{\mathrm{ss}}^{(\alpha)}]=\lambda{\cal L}_{\alpha}[\rho]+(1-\lambda){\cal L}_{\alpha}[\rho_{\mathrm{ss}}^{(\alpha)}]$.
And remember we require ${\cal L}_{\alpha}[\rho_{\mathrm{ss}}^{(\alpha)}]=0$,
thus, the above inequality becomes 
\begin{equation}
-\lambda\mathrm{tr}\left[{\cal L}_{\alpha}[\rho]\cdot\ln[\lambda\rho+(1-\lambda)\rho_{\mathrm{ss}}^{(\alpha)}]\right]\le-\lambda\mathrm{tr}\left[{\cal L}_{\alpha}[\rho]\cdot\ln\rho\right].
\end{equation}
 In the limit $\lambda\rightarrow0^{+}$, we obtain
\begin{equation}
\mathrm{tr}\left[{\cal L}_{\alpha}[\rho]\cdot(\ln\rho_{\mathrm{ss}}^{(\alpha)}-\ln\rho)\right]=R_{\mathrm{Sp}}^{(\alpha)}\ge0.
\end{equation}
 Therefore, we have $R_{\mathrm{Sp}}=\sum R_{\mathrm{Sp}}^{(\alpha)}\ge0$.
\quad  $\blacksquare$

\section{Properties of a squeezed bath and the master equation \label{apx:squeeze}}

\textbf{1. Squeezed bath properties} - Here we show some basic properties
of a squeezed thermal bath. The Hamiltonian of the bath is $\hat{H}_{B}=\sum_{k}\omega_{k}\hat{b}_{k}^{\dagger}\hat{b}_{k}$,
and the squeezed thermal state is
\begin{equation}
\rho_{B}:={\cal S}\cdot\rho_{\mathrm{th}}\cdot{\cal S}^{\dagger},\qquad\rho_{\mathrm{th}}:=\frac{1}{{\cal Z}}\exp[-\beta\,\hat{H}_{B}].
\end{equation}
Here ${\cal S}$ is the squeezing operator for the boson bath,
\begin{equation}
{\cal S}:=\prod_{k}\mathfrak{s}_{k}(\lambda_{k}),\qquad\mathfrak{s}_{k}(\lambda_{k}):=\exp[\frac{1}{2}\lambda_{k}^{*}\hat{b}_{k}^{2}-\mathbf{h.c.}],\qquad\lambda_{k}:=r_{k}e^{i\theta_{k}}\,(r_{k}>0),
\end{equation}
and $\mathfrak{s}_{k}$ is the squeezing operator for mode $\hat{b}_{k}$
in the bath. They satisfy
\begin{align}
\mathfrak{s}_{k}^{\dagger}(\lambda_{k})\hat{b}_{k}\mathfrak{s}_{k}(\lambda_{k}) & =\hat{b}_{k}+[\frac{1}{2}\big(\lambda_{k}(\hat{b}_{k}^{\dagger})^{2}-\lambda_{k}^{*}\hat{b}_{k}^{2}\big),\,\hat{b}_{k}]+\frac{1}{2!}[\frac{1}{2}\big(\lambda_{k}(\hat{b}_{k}^{\dagger})^{2}-\lambda_{k}^{*}\hat{b}_{k}^{2}\big),\,[\frac{1}{2}\big(\lambda_{k}(\hat{b}_{k}^{\dagger})^{2}-\lambda_{k}^{*}\hat{b}_{k}^{2}\big),\,\hat{b}_{k}]]+\dots\nonumber \\
 & =\hat{b}_{k}-\lambda_{k}\hat{b}_{k}^{\dagger}+\frac{1}{2!}|\lambda_{k}|^{2}\hat{b}_{k}-\frac{1}{3!}\lambda_{k}|\lambda_{k}|^{2}\hat{b}_{k}^{\dagger}-\frac{1}{4!}|\lambda_{k}|^{4}\hat{b}_{k}+\dots=\hat{b}_{k}\cosh r_{k}-\hat{b}_{k}^{\dagger}e^{i\theta_{k}}\sinh r_{k},\\
\mathfrak{s}_{k}(\lambda_{k})\hat{b}_{k}\mathfrak{s}_{k}^{\dagger}(\lambda_{k}) & =\hat{b}_{k}\cosh r_{k}+\hat{b}_{k}^{\dagger}e^{i\theta_{k}}\sinh r_{k}.\nonumber 
\end{align}
Thus we have
\begin{align}
\tilde{\mathsf{u}}_{k}: & =-\mathrm{tr}[\rho_{B}\cdot\hat{b}_{k}^{2}]=-\mathrm{tr}[\rho_{\mathrm{th}}\cdot\mathfrak{s}_{k}^{\dagger}\hat{b}_{k}\mathfrak{s}_{k}\cdot\mathfrak{s}_{k}^{\dagger}\hat{b}_{k}\mathfrak{s}_{k}]=-\mathrm{tr}[\rho_{\mathrm{th}}\cdot(\hat{b}_{k}\cosh r_{k}-\hat{b}_{k}^{\dagger}e^{i\theta_{k}}\sinh r_{k})\cdot(\hat{b}_{k}\cosh r_{k}-\hat{b}_{k}^{\dagger}e^{i\theta_{k}}\sinh r_{k})]\nonumber \\
 & =\cosh r_{k}\sinh r_{k}e^{i\theta_{k}}(2\overline{\mathrm{n}}_{k}+1)=e^{i\theta_{k}}\sinh2r_{k}\,(\overline{\mathrm{n}}_{k}+\frac{1}{2}),\\
\tilde{\mathsf{n}}_{k}: & =\mathrm{tr}[\rho_{B}\cdot\hat{b}_{k}^{\dagger}\hat{b}_{k}]=\mathrm{tr}[\rho_{\mathrm{th}}\cdot\mathfrak{s}_{k}^{\dagger}\hat{b}_{k}^{\dagger}\mathfrak{s}_{k}\cdot\mathfrak{s}_{k}^{\dagger}\hat{b}_{k}\mathfrak{s}_{k}]=\mathrm{tr}[\rho_{\mathrm{th}}\cdot(\hat{b}_{k}^{\dagger}\cosh r_{k}-\hat{b}_{k}e^{-i\theta_{k}}\sinh r_{k})\cdot(\hat{b}_{k}\cosh r_{k}-\hat{b}_{k}^{\dagger}e^{i\theta_{k}}\sinh r_{k})]\nonumber \\
 & =\cosh^{2}r_{k}\cdot\overline{\mathrm{n}}_{k}+\sinh^{2}r_{k}\cdot(\overline{\mathrm{n}}_{k}+1)=\cosh2r_{k}\,(\overline{\mathrm{n}}_{k}+\frac{1}{2})-\frac{1}{2},
\end{align}
where $\overline{\mathrm{n}}_{k}:=[\exp(\beta\omega_{k})-1]^{-1}$
is the Planck distribution. 

\vspace{.5cm}

\textbf{2. Master equation derivation} - Now we derive the master
equation for a single boson mode ($\hat{H}_{S}=\Omega\hat{a}^{\dagger}\hat{a}$)
interacting with a squeezed boson bath. The interaction Hamiltonian
is $\hat{V}_{SB}=\hat{a}\hat{B}^{\dagger}+\hat{a}^{\dagger}\hat{B}$,
where $\hat{B}=\sum_{k}g_{k}\hat{b}_{k}$, and the master equation
is derived by
\begin{align}
\dot{\rho} & =-\mathrm{tr}_{B}\int_{0}^{\infty}ds\,[\tilde{V}_{SB}(t-s),\,[\tilde{V}_{SB}(t),\,\rho(t)\otimes\rho_{B}]]\nonumber \\
 & =\mathrm{tr}_{B}\int_{0}^{\infty}ds\,[\tilde{V}_{SB}(t-s)\rho(t)\rho_{B}\tilde{V}_{SB}(t)-\tilde{V}_{SB}(t-s)\tilde{V}_{SB}(t)\rho(t)\rho_{B}]+\mathbf{h.c.}
\end{align}
Here we use $\tilde{o}(t)$ to denote the operators in the interaction
picture, and $\tilde{a}(t)=\hat{a}e^{-i\Omega t}$, $\tilde{b}_{k}(t)=\hat{b}_{k}e^{-i\omega_{k}t}$.
We adopted the Born approximation $\rho_{SB}(t)\simeq\rho(t)\otimes\rho_{B}$,
and
\begin{align}
\rho_{B}(t) & \simeq\rho_{B}^{0}=\frac{1}{{\cal Z}}\,\exp[-\beta\,{\cal S}\hat{H}_{B}{\cal S}^{\dagger}].
\end{align}
We define the coupling spectrum as
\begin{equation}
J(\omega):=\sum_{k}|g_{k}|^{2}\cdot\delta(\omega-\omega_{k}),\qquad K(\omega):=\sum_{k}g_{k}^{2}\cdot\delta(\omega-\omega_{k}).
\end{equation}
We omit the phase of $g_{k}$, thus we have $K(\omega)=J(\omega)=K^{*}(\omega)$.
Here is the calculation for some terms:
\begin{align}
\int_{0}^{\infty}ds\, & \mathrm{tr}_{B}\left[\tilde{a}^{\dagger}(t-s)\tilde{B}(t-s)\cdot\rho(t)\rho_{B}\cdot\tilde{a}(t)\tilde{B}^{\dagger}(t)\right]=\int_{0}^{\infty}ds\,\hat{a}^{\dagger}\rho\hat{a}e^{-i\Omega s}\cdot\mathrm{tr}_{B}[\rho_{B}\tilde{B}^{\dagger}(t)\tilde{B}(t-s)]\nonumber \\
= & \hat{a}^{\dagger}\rho\hat{a}\int_{0}^{\infty}\frac{d\omega}{2\pi}\int_{0}^{\infty}ds\,e^{-i\Omega s}\cdot e^{i\omega s}\,J(\omega)\tilde{\mathsf{n}}(\omega)=\frac{1}{2}\gamma\tilde{\mathfrak{n}}\cdot\hat{a}^{\dagger}\rho\hat{a},
\end{align}
\begin{align}
\int_{0}^{\infty}ds\, & \mathrm{tr}_{B}\left[\tilde{a}^{\dagger}(t-s)\tilde{B}(t-s)\cdot\rho(t)\rho_{B}\cdot\tilde{a}^{\dagger}(t)\tilde{B}(t)\right]=\int_{0}^{\infty}ds\,\hat{a}^{\dagger}\rho\hat{a}^{\dagger}e^{2i\Omega t}\cdot e^{-i\Omega s}\cdot\mathrm{tr}_{B}[\rho_{B}\tilde{B}(t)\tilde{B}(t-s)]\nonumber \\
= & -\hat{a}^{\dagger}\rho\hat{a}^{\dagger}e^{2i\Omega t}\int_{0}^{\infty}\frac{d\omega}{2\pi}\int_{0}^{\infty}ds\,e^{-i\Omega s}\cdot e^{i\omega s}\cdot e^{-2i\omega t}\,K(\omega)\tilde{\mathsf{u}}(\omega)=-\frac{1}{2}\gamma\tilde{\mathfrak{u}}\cdot\hat{a}^{\dagger}\rho\hat{a}^{\dagger},
\end{align}
where $\gamma=J(\Omega)=K(\Omega)$ and
\begin{equation}
\tilde{\mathfrak{n}}=\cosh2r_{\Omega}\,(\overline{\mathrm{n}}_{\Omega}+\frac{1}{2})-\frac{1}{2},\qquad\tilde{\mathfrak{u}}=e^{i\theta_{\Omega}}\sinh2r_{\Omega}\,(\overline{\mathrm{n}}_{\Omega}+\frac{1}{2}).
\end{equation}
We omitted all the Principal integral in the above calculation. Thus
the master equation is (interaction picture)
\begin{align}
\dot{\rho}= & \gamma\tilde{\mathfrak{n}}\big(\hat{a}^{\dagger}\rho\hat{a}-\frac{1}{2}\{\hat{a}\hat{a}^{\dagger},\rho\}\big)+\gamma(\tilde{\mathfrak{n}}+1)\big(\hat{a}\rho\hat{a}^{\dagger}-\frac{1}{2}\{\hat{a}^{\dagger}\hat{a},\rho\}\big)\nonumber \\
 & -\gamma\tilde{\mathfrak{u}}\big(\hat{a}^{\dagger}\rho\hat{a}^{\dagger}-\frac{1}{2}\{(\hat{a}^{\dagger})^{2},\rho\}\big)+\gamma\tilde{\mathfrak{u}}^{*}\big(\hat{a}\rho\hat{a}-\frac{1}{2}\{(\hat{a})^{2},\rho\}\big).\label{eq:ME}
\end{align}

From the above master equation, we obtain 
\begin{equation}
\frac{d}{dt}\langle\tilde{a}(t)\rangle=-\frac{\gamma}{2}\langle\tilde{a}\rangle,\qquad\frac{d}{dt}\langle\tilde{a}^{\dagger}\tilde{a}\rangle=-\gamma[\langle\tilde{n}_{a}\rangle-\tilde{\mathfrak{n}}],\qquad\frac{d}{dt}\langle\tilde{a}^{2}\rangle=-\gamma[\langle\tilde{a}^{2}\rangle-\tilde{\mathfrak{u}}].
\end{equation}
In the steady state we have $\langle\tilde{a}\rangle_{\mathrm{ss}}=0$,
$\langle\tilde{a}^{\dagger}\tilde{a}\rangle_{\mathrm{ss}}=\tilde{\mathfrak{n}}$
and $\langle\tilde{a}^{2}\rangle_{\mathrm{ss}}=\tilde{\mathfrak{u}}$.
Thus we can verify the steady state is
\begin{equation}
\rho_{\mathrm{ss}}=\frac{1}{Z}\exp[-\beta\Omega\,\mathsf{S}\hat{a}^{\dagger}\hat{a}\mathsf{S}^{\dagger}],\qquad\mathsf{S}=\exp[-\frac{1}{2}\zeta^{*}\hat{a}^{2}+\frac{1}{2}\zeta(\hat{a}^{\dagger})^{2}],\qquad\zeta:=\lambda_{k}\big|_{\omega_{k}=\Omega}.
\end{equation}
Here $\mathsf{S}$ is a squeezing operator for the system, and we
remark that the above $\rho_{\mathrm{ss}}$ is in the interaction
picture. When the single boson is coupled with multiple squeezed baths,
the generalization is straightforward, as shown in the main text. 

\vspace{.5cm}

\textbf{3. Time correlation functions }- From the above equations
of $\langle\tilde{a}(t)\rangle$, we obtain $\langle\tilde{a}(t)\rangle=\langle\tilde{a}(s)\rangle e^{-\frac{\gamma}{2}(t-s)}$
($t\ge s$). According to the quantum regression theorem, we know
the time correlation functions satisfy the following equations ($t\ge s$)
\cite{breuer_theory_2002,gardiner_quantum_2004}
\begin{equation}
\frac{d}{dt}\langle\tilde{a}^{\dagger}(t)\tilde{a}(s)\rangle=-\frac{\gamma}{2}\langle\tilde{a}^{\dagger}(t)\tilde{a}(s)\rangle,\qquad\frac{d}{dt}\langle\tilde{a}(t)\tilde{a}(s)\rangle=-\frac{\gamma}{2}\langle\tilde{a}(t)\tilde{a}(s)\rangle.
\end{equation}
Here $\langle\tilde{o}_{1}(t)\tilde{o}_{2}(s)\rangle$ are correlation
functions in the rotating frame, defined by $\langle\tilde{o}_{1}(t)\tilde{o}_{2}(s)\rangle=\mathrm{tr}[\hat{o}_{1}\,{\cal E}_{t-s}\hat{o}_{2}\,{\cal E}_{s}\rho(0)]$
for $t\ge s$ \cite{breuer_theory_2002}, where $\hat{o}_{1,2}$ are
operators in Schr\"odinger picture, and ${\cal E}_{t}$ is the evolution
operator solved from the above master equation in interaction picture
{[}Eq.\,(\ref{eq:ME}){]}, and $\rho(t)={\cal E}_{t-s}\rho(s)$.
Similarly, $\langle\hat{o}_{1}(t)\hat{o}_{2}(s)\rangle$ are correlation
functions without adopting the rotating frame, and we have
\begin{align}
\langle\hat{a}^{\dagger}(t)\hat{a}(s)\rangle & =\langle\tilde{a}^{\dagger}(t)\tilde{a}(s)\rangle e^{i\Omega(t-s)}=\langle\tilde{a}^{\dagger}(s)\tilde{a}(s)\rangle e^{i\Omega(t-s)}\cdot e^{-\frac{\gamma}{2}(t-s)},\nonumber \\
\langle\hat{a}(t)\hat{a}(s)\rangle & =\langle\tilde{a}(t)\tilde{a}(s)\rangle e^{-i\Omega(t+s)}=\langle\tilde{a}^{2}(s)\rangle e^{-2i\Omega s}\cdot e^{-i\Omega(t-s)}\cdot e^{-\frac{\gamma}{2}(t-s)}.
\end{align}

This can be also calculated using the Langevin equation $\dot{\hat{a}}=-i\Omega\hat{a}-\frac{1}{2}\gamma\hat{a}-\hat{\xi}(t)$
(here we only consider one single bath), e.g.,
\begin{align*}
\langle\hat{a}^{\dagger}(t)\hat{a}(s)\rangle & =\Big\langle[\hat{a}^{\dagger}(s)e^{(i\Omega-\frac{\gamma}{2})(t-s)}-\int_{s}^{t}dt'\,e^{(i\Omega-\frac{\gamma}{2})(t-t')}\hat{\xi}^{\dagger}(t')]\cdot\hat{a}(s)\Big\rangle\\
 & =\langle\hat{a}^{\dagger}(s)\hat{a}(s)\rangle e^{(i\Omega-\frac{\gamma}{2})(t-s)}-\int_{s}^{t}dt'\,e^{(i\Omega-\frac{\gamma}{2})(t-t')}\Big\langle\hat{\xi}^{\dagger}(t')\cdot[\hat{a}(0)e^{(-i\Omega-\frac{\gamma}{2})s}-\int_{0}^{s}ds'\,e^{(-i\Omega-\frac{\gamma}{2})(s-s')}\hat{\xi}(s')]\Big\rangle\\
 & =\langle\hat{a}^{\dagger}(s)\hat{a}(s)\rangle e^{(i\Omega-\frac{\gamma}{2})(t-s)}+\int_{s}^{t}dt'\int_{0}^{s}ds'\,e^{(i\Omega-\frac{\gamma}{2})(t-t')}\,e^{(-i\Omega-\frac{\gamma}{2})(s-s')}\langle\hat{\xi}^{\dagger}(t')\hat{\xi}(s')\rangle.
\end{align*}
Under the Markovian approximation we have $\langle\hat{\xi}^{\dagger}(t')\hat{\xi}(s')\rangle\sim\delta(t'-s')$
\cite{scully_quantum_1997,gardiner_quantum_2004}. And notice that
in the above double integral we have $0\le s'\le s\le t'\le t$, thus
the above integral gives zero. 

\vspace{.5cm}

\textbf{4. Bath entropy change }- Now we show the calculation for
the 2nd relation of Eq.\,(\ref{eq:F-k}) in the main text. Using
the Heisenberg equation we obtain
\begin{equation}
\frac{d}{dt}\sum_{k}\mathfrak{h}_{k}\langle\tilde{b}_{\alpha k}^{2}(t)\rangle+\mathbf{h.c.}=\sum_{k}-i2g_{\alpha k}^{*}\mathfrak{h}_{k}\Big[\langle\hat{a}(t)\hat{b}_{\alpha k}(0)\rangle e^{i\omega_{\alpha k}t}-ig_{\alpha k}^{*}\int_{0}^{t}ds\,e^{i\omega_{\alpha k}(t+s)}\langle\hat{a}(t)\hat{a}(s)\rangle\Big]+\mathbf{h.c.}\label{eq:part2}
\end{equation}
The 1st term in the bracket is further calculated by substituting
$\hat{a}(t)$ {[}Eq.\,(\ref{eq:a(t)}){]}, and it gives
\begin{align}
 & -\sum_{k}2|g_{\alpha k}|^{2}\mathfrak{h}_{k}\int_{0}^{t}ds\,e^{-[i(\Omega-\omega_{\alpha k})+\frac{\Gamma}{2}](t-s)}\langle\hat{b}_{\alpha k}^{2}(0)\rangle+\mathbf{h.c.}\nonumber \\
\simeq & 2\int_{0}^{\infty}\frac{d\omega}{2\pi}\int_{0}^{\infty}ds\,J_{\alpha}(\omega)\mathfrak{h}(\omega)e^{-i(\Omega-\omega)s-\frac{1}{2}\Gamma s}\tilde{\mathsf{u}}_{\alpha}(\omega)+\mathbf{h.c.}\nonumber \\
= & 2\int_{0}^{\infty}\frac{d\omega}{2\pi}\,J_{\alpha}(\omega)[\frac{\mathfrak{h}(\omega)\tilde{\mathsf{u}}_{\alpha}(\omega)}{\frac{\Gamma}{2}+i(\Omega-\omega)}+\mathbf{h.c.}]\simeq\gamma_{\alpha}[\mathfrak{h}(\Omega)\tilde{\mathfrak{u}}_{\alpha}+\mathbf{h.c.}].\label{eq:part2-1}
\end{align}
Here we applied the Markovian approximation and the weak coupling
limit $\Gamma\ll\Omega$. The 2nd term of Eq.\,(\ref{eq:part2})
can be calculated with the help of the relation (quantum regression
theorem)
\begin{align}
\langle\hat{a}(t)\hat{a}(s)\rangle & =\langle\tilde{a}^{2}(s)\rangle e^{-2i\Omega s}\cdot e^{-i\Omega(t-s)-\frac{\Gamma}{2}(t-s)},\quad\text{for }t\ge s
\end{align}
and it leads to
\begin{align}
 & -\sum_{k}2(g_{\alpha k}^{*})^{2}\mathfrak{h}_{k}e^{2i\omega_{\alpha k}t}\int_{0}^{t}ds\,e^{-i\omega_{\alpha k}(t-s)}\langle\hat{a}(t)\hat{a}(s)\rangle+\mathbf{h.c.}\nonumber \\
\simeq & -2\int\frac{d\omega}{2\pi}K_{\alpha}(\omega)\mathfrak{h}(\omega)e^{2i(\omega-\Omega)t}\langle\tilde{a}^{2}(t)\rangle\int_{0}^{\infty}ds\,e^{[i(\Omega-\omega)-\frac{\Gamma}{2}]s}+\mathbf{h.c.}\nonumber \\
= & -2\langle\tilde{a}^{2}(t)\rangle\cdot\int_{0}^{\infty}\frac{d\omega}{2\pi}[\frac{K_{\alpha}(\omega)\mathfrak{h}(\omega)e^{2i(\omega-\Omega)t}}{\frac{\Gamma}{2}+i(\omega-\Omega)}+\mathbf{h.c.}]\simeq-\gamma_{\alpha}[\mathfrak{h}(\Omega)\langle\tilde{a}^{2}(t)\rangle+\mathbf{h.c.}].\label{eq:part2-2}
\end{align}

Thus, summing up Eqs.\,(\ref{eq:part2-1}, \ref{eq:part2-2}), we
finish our calculation 
\begin{equation}
\frac{d}{dt}\sum_{k}\mathfrak{h}_{k}\langle\tilde{b}_{\alpha k}^{2}(t)\rangle+\mathbf{h.c.}=-\mathfrak{h}(\Omega)\cdot\gamma_{\alpha}[\langle\tilde{a}^{2}(t)\rangle-\tilde{\mathfrak{u}}_{\alpha}]+\mathbf{h.c.}
\end{equation}

\end{widetext}

\bibliographystyle{apsrev4-1}
\bibliography{Refs}

\end{document}